\documentclass[12pt]{article}
\usepackage{latexsym} \usepackage{epsf}
\usepackage{epsfig}
\usepackage{a4}
\usepackage{amsfonts}
\usepackage{color}
\newcommand{\ba}{\begin{array}}
\newcommand{\ea}{\end{array}}
\newcommand{\be}{\begin{equation}}
\newcommand{\ee}{\end{equation}}
\newcommand{\nn}{\nonumber}
\newcommand{\bea}{\begin{eqnarray}}
\newcommand{\ena}{\end{eqnarray}}
\newcommand{\beas}{\begin{eqnarray*}}
\newcommand{\enas}{\end{eqnarray*}}

\begin{document}
\begin{center}
 {\Large{Anizotropic Ising Model on 2D Kagom\'{e}  Lattice as an Inhomogeneous XYZ Integrable Model }}
\end{center}

\begin{center}
	{\bf Shahane A. Khachatryan{\footnote{e-mail:{\sl shah@mail.yerphi.am}}$^{,a}$},
		Hrachya Babujian{\footnote{e-mail:{\sl babujian@yerphi.am}}$^{,a,b}$  and
			Ara G. Sedrakyan
			{\footnote{e-mail:{\sl sedrak@mail.yerphi.am}}}$^{,a}$}}\\
	
	\vspace{12pt}
	
	{\it $^a$ Alikhanyan National Sience Laboratory (Yerevan Physics Institute), \\Alikhanian Br. str. 2, Yerevan 36,  Armenia}\\
	\vspace{6pt}
	$^b$ Beijing Institute of Mathematical Sciences and Applications,
	\\Huairou 101408, Beijing, China
\end{center}


\begin{abstract}
	We investigate a generalized inhomogeneous two-dimensional Ising model on the kagom\'e lattice with two alternating couplings $J_i$ and $J'_i$, $i=1,2,3$, along the three lattice directions. The local Boltzmann weights are mapped onto a non-symmetric eight-vertex $R$-matrix satisfying the free-fermion condition for arbitrary values of the six couplings. We analyze the Yang--Baxter structure for the cases $J'_i=J_i$ and $J'_i=-J_i$ and derive the corresponding one-dimensional quantum spin chains in the anisotropic limit. The first case yields a transverse-field Ising-type Hamiltonian, while the second leads to a modified chain with a graded permutation structure and shifted partition-function zeros. For a partially anisotropic model, we also obtain the free energy and specific heat.
\end{abstract}

%

\section{Introduction}

The  two-dimensional Ising model (2DIM) is one of the fundamental exactly solvable models of statistical physics and provides a basic framework for understanding phase transitions, critical phenomena, and the relation between classical statistical systems and quantum many-body models. While the square, triangular, and honeycomb lattices represent the simplest regular realizations of the 2DIM, the kagom\'{e} lattice is of particular interest because of its nontrivial geometry, formed by corner-sharing triangles and hexagons, and because of the frustration effects appearing in the antiferromagnetic regime.

The Ising model on the kagom\'{e} lattice has a long history. Its thermodynamic and critical properties were investigated already in the early works on the kagom\'{e} Ising net \cite{Kagome1}. Further studies addressed spontaneous magnetization, ferrimagnetic properties, symmetry relations, and complex-temperature singularities of the kagom\'{e} and related planar Ising models \cite{Kagome2}. The geometrical frustration inherent to lattices containing triangular plaquettes gives rise to a particularly rich behavior in the antiferromagnetic case and distinguishes the kagom\'{e} lattice from unfrustrated two-dimensional lattices.

More generally, exact and systematic methods for Ising models on planar lattices have attracted renewed interest. The Feynman-Vdovichenko combinatorial approach was used in \cite{Codello-2010} to determine exact Curie temperatures for all Archimedean lattices and, by duality, for the corresponding Laves lattices. The mathematical structure of the Kac--Ward formulation and its relation to critical Ising models, duality, and discrete Laplacians was developed in \cite{Cimasoni-2012}. More recently, a universal-emulator formulation based on the Feynman--Vdovichenko/Kac--Ward construction has been proposed for a broad class of planar Ising lattices, including Archimedean and $2$-uniform lattices \cite{Codello-2026}. The effects of geometrical frustration in Ising antiferromagnets on Archimedean lattices, including the kagom\'{e} lattice, were systematically investigated in \cite{Yu-2015}, while their nonequilibrium zero-temperature Glauber dynamics was studied in Ref.~\cite{Yu-2017}. General expressions for the free energy, entropy, and specific heat of planar Ising models and their application to Archimedean lattices and their duals were obtained in Ref.~\cite{Pierrel-2025}. Very recently, exact bounds on the critical temperature of two-dimensional periodic Ising lattices were derived in Ref.~\cite{Joseph-2026-2}, and families of planar lattices with arbitrarily high critical temperatures were constructed in Ref.~\cite{Joseph-2026-1}. These developments demonstrate that the relation between lattice geometry, fermionic formulations, and critical properties of planar Ising models continues to be an active subject.

An important alternative formulation of the 2DIM is based on its relation to vertex models. The local Boltzmann weights of the Ising model can be transformed into an $R$-matrix of the eight-vertex type. This representation establishes a direct connection between a classical two-dimensional statistical model and one-dimensional quantum spin chains \cite{Baxter1,Baxter,Suz,Cardy} and, at the same time, makes it possible to investigate the Yang--Baxter equations \cite{YangOns} and the integrability properties of the corresponding model. In Ref.~\cite{KhS1}, such an approach was developed for the 2DIM on the regular lattice. The partition function was reformulated as a trace of a product of local $R$-matrices, and the relation between the eight-vertex and fermionic representations was established.

A further direction, particularly relevant for the present work, is based on the relation between two-dimensional classical statistical models, vertex models, and quantum integrable systems. In Ref.~\cite{KhS1}, we developed a fermionic field-theoretical formulation of the two-dimensional Ising and XYZ models, in which the local Boltzmann weights are represented in terms of $R$-matrices. The free-fermion structure of the corresponding $R$-matrix makes it possible to formulate the partition function and correlation functions in terms of fermionic variables and establishes a direct connection between the classical 2D Ising model and vertex models. In particular, determinant representations for the spin-spin correlation functions were obtained within this approach.
	
The integrability properties of a more general class of vertex models were subsequently investigated in Ref.~\cite{KhS2}, where the Yang--Baxter equations for general inhomogeneous six- and eight-vertex $R$-matrices were analyzed. The spectral-parameter-dependent solutions were classified under the general condition of particle-number conservation modulo two, and both one-parameter and non-homogeneous two-parameter solutions of the Yang--Baxter equations were obtained. Their relation to Zamolodchikov's tetrahedral algebra \cite{Zamolod} was also investigated. These results provide the algebraic framework for the $R$-matrix construction used in the present work and allow us to investigate the integrability of the generalized inhomogeneous Kagom'{e} Ising model and the corresponding one-dimensional quantum spin chains.

The present work is the third step in a sequence of our recent investigations aimed at developing complementary exact formulations of the inhomogeneous kagom\'{e} Ising model.

In the first work of this sequence, Ref.~\cite{FerKagome}, we investigated the inhomogeneous 2DIM on the kagom\'{e} lattice by mapping it onto a particular non-symmetric eight-vertex model and constructing its fermionic representation. The partition function was reduced to a Grassmann functional integral with a quadratic fermionic action, which made it possible to obtain the free energy and the exact critical surface in the thermodynamic limit. The specific heat and spontaneous magnetization were also investigated in the ferromagnetic case. As important consistency checks, it was shown that when one or two coupling constants vanish, the model reduces, respectively, to the square-lattice and one-dimensional Ising models.

In the second work, Ref.~\cite{KWK}, we developed a complementary fermionic description based directly on the Kac--Ward representation. In this approach, Grassmann variables are associated with directed lattice links, while the geometrical turning of a fermionic trajectory is encoded in the corresponding Kac--Ward phase factors. The partition function is represented through the determinant of a finite-dimensional matrix in momentum space, and the zeros of this determinant determine the excitation spectrum and the critical couplings. The same construction was applied to the square, honeycomb, triangular, kagom\'{e}, and dual kagom\'{e} lattices. In particular, for the anisotropic kagom\'{e} lattice the complete critical surface obtained from the fermionic determinant coincides with that found independently in Ref.~\cite{FerKagome}. Thus, the first two works provide two complementary fermionic derivations of the thermodynamic and critical properties of the kagom\'{e} Ising model.

In the present, third work of this sequence, we concentrate on the $R$-matrix and integrability aspects of the same problem. We consider a generalized inhomogeneous kagom\'{e} Ising model in which two alternating coupling constants $J_i$ and $J_i'$, $i=1,2,3$, are assigned to each of the three lattice directions. By combining two neighbouring triangles into an elementary checkerboard cell, we construct the corresponding local Boltzmann-weight matrix. After a local unitary transformation, this matrix takes the form of a non-symmetric eight-vertex $R$-matrix. An important property of the resulting $R$-matrix is that it satisfies the free-fermion condition for arbitrary values of the six coupling constants $J_i$ and $J_i'$.

We then investigate the Yang--Baxter equations associated with this $R$-matrix and analyze the corresponding integrability conditions. Particular attention is paid to the two choices $J_i'=J_i$ and $J_i'=-J_i$. The first case contains the ordinary anisotropic kagom\'{e} Ising model, while the second defines a sign-altered model with a different structure of the corresponding $R$-matrix. For a partly anisotropic choice of the coupling constants, we also present the corresponding expression for the free energy.

Finally, using the anisotropic, or continuum, limit of the transfer matrix, we derive the one-dimensional quantum spin models associated with the classical two-dimensional kagom\'{e} Ising model. For $J_i'=J_i$, the resulting quantum Hamiltonian is of the transverse-field Ising type, and its critical point agrees with the corresponding limit of the critical surface of the classical kagom\'{e} model. For the sign-altered case $J_i'=-J_i$, the limiting $R$-matrix involves a graded permutation matrix and leads to a modified one-dimensional Ising Hamiltonian. In this case, the zeros of the partition function occur at shifted momenta, in contrast to the conventional ferromagnetic case.

Thus, the three works provide mutually complementary descriptions of the inhomogeneous kagom\'{e} Ising model: the first develops the explicit fermionic-field solution and thermodynamics, the second formulates criticality and the low-energy spectrum within the Kac--Ward approach, and the present work develops the $R$-matrix, Yang--Baxter, and quantum-spin-chain formulation. Taken together, these approaches establish a direct connection between the geometry of the kagom\'{e} lattice, its fermionic representation, exact critical properties, and the integrable structure of the associated quantum models.

\section{The partition function: $R$-matrix}

The kagom'{e} lattice is shown in Fig.~1. It consists of periodically arranged triangles and hexagons. The Ising spins $s_\alpha=\pm1$ are located at the vertices of the lattice, and only nearest-neighbour interactions are considered. In the inhomogeneous case, the spin-spin interactions are characterized by three different coupling constants $J_i$, $i=1,2,3$, associated with the links oriented along the three different lattice directions. In this work, we consider a more general version of the model, with two alternating coupling constants $J_i$ and $J_i'$ assigned to each direction $i=1,2,3$, as shown in Fig.~1. The partition function can then be written as
 \bea
 Z=\sum_{s}\prod_{\alpha,\beta,\gamma,\alpha',\beta',\gamma'}
 e^{J_1' s_{\alpha}s_{\beta'}+
 J_1 s_{\beta}s_{\alpha'}+J_2 s_\beta s_\gamma+J_2' s_\gamma s_{\beta'}+J_3 s_\alpha s_\gamma+J'_3 s_\gamma s_{\alpha'}}.
 \ena
 %
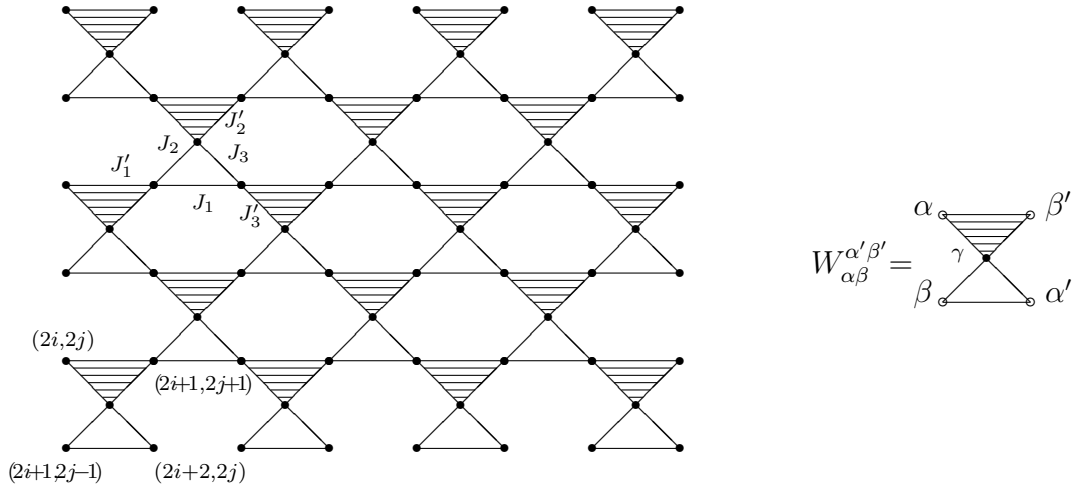
\begin{figure}[h]
	\unitlength=11pt
	\begin{picture}(100,15)(-3,-1)
		
		\newsavebox{\rw}
		
		\sbox{\rw}{\begin{picture}(3,3)
				\put(0,0){\line(1,1){3}}\put(0,3){\line(1,-1){3}}
				\put(1,2){\line(1,0){1}}\put(1.25,1.75){\line(1,0){0.5}}
				\put(0.5,2.5){\line(1,0){2}}\put(0.75,2.25){\line(1,0){1.5}}\put(0.25,2.75){\line(1,0){2.5}}
				\put(0,0){\line(1,0){3}}\put(0,3){\line(1,0){3}}\put(1.5,1.5){\circle*{0.25}}
				\put(0,0){\circle*{0.25}}\put(0,3){\circle*{0.25}}\put(3,0){\circle*{0.25}}\put(3,3){\circle*{0.25}}
		\end{picture}}
		\multiput(0,-1)(0,6){3}{\usebox{\rw}}\multiput(6,-1)(0,6){3}{\usebox{\rw}}
		\multiput(12,-1)(0,6){3}{\usebox{\rw}}\multiput(18,-1)(0,6){3}{\usebox{\rw}}
		\multiput(3,2)(0,6){2}{\usebox{\rw}}\multiput(9,2)(0,6){2}{\usebox{\rw}}
		\multiput(15,2)(0,6){2}{\usebox{\rw}}\put(1.5,8.5){\scriptsize${J_1'}$}\put(4.3,7.2){\scriptsize${J_1}$}
		\put(3.1,9.2){\scriptsize${J_2}$}\put(5.4,10){\scriptsize${J_2'}$}\put(5.5,8.9){\scriptsize${J_3}$}
		\put(5.9,6.8){\scriptsize${J_3'}$}
		\put(-1.2,2.5){$\scriptstyle(2i,2j)$}\put(-2,-2){$\scriptstyle(\!2i\!+\!1,\!2j\!-\!1\!)$}
		\put(3,1){$\scriptstyle(\!2i\!+\!1,2j\!+\!1\!)$}\put(3,-2){$\scriptstyle(2i+2,2j)$}
		\put(25.5,5){${W}_{\alpha\beta}^{\alpha'\beta'}$=}
		\put(30,4){\line(1,1){3}}\put(30,7){\line(1,-1){3}}
		\put(30,4){\line(1,0){3}}\put(30,7){\line(1,0){3}}\put(30.25,6.75){\line(1,0){2.5}}
		\put(30.5,6.5){\line(1,0){2}}\put(30.75,6.25){\line(1,0){1.5}}\put(31,6){\line(1,0){1}}
		\put(31.25,5.75){\line(1,0){0.5}}
		\put(31.5,5.5){\circle*{0.25}}
		\put(29,4){$\beta$}\put(29,7){$\alpha$}\put(33.5,4){$\alpha'$}
		\put(33.5,7){$\beta'$}
		\put(30.3,5.5){\scriptsize$\gamma$}
		\put(30,4){\circle{0.25}}\put(30,7){\circle{0.25}}\put(33,4){\circle{0.25}}\put(33,7){\circle{0.25}}
	\end{picture} \caption{Kagom\'{e} lattice}\label{fig1}
\end{figure}

However, it is possible to represent the lattice in a checkerboard form with square cells, each constructed from two triangles, as shown in Fig.~\ref{fig1} by $W_{ij}$. In this representation, the partition function can be rewritten as
\bea
Z=\sum_{{s_{\alpha\beta}}}\prod_{\alpha,\beta,\alpha',\beta'}W_{\alpha\beta}^{\alpha' \beta'},\quad W_{\alpha\beta}^{\alpha' \beta'}=\sum_{s_\gamma=\pm 1}e^{J_1' s_{\alpha}s_{\beta'}+J_1 s_{\beta}s_{\alpha'}+J_2 s_\beta  s_\gamma+J_2' s_{\beta'} s_\gamma +J_3 s_\alpha  s_\gamma+J'*3 s*\gamma s_{\alpha'}  }.
\label{partw}
\ena
We see that the weight function $W_{\alpha\beta}$ is associated with two neighbouring triangles sharing a common vertex. There are three different ways of choosing such pairs of triangles. For a large lattice, either infinite or with periodic boundary conditions, the partition function is invariant under such a choice, 
which corresponds to an interchange of the parameters $J_i,\; J_i'$, $i=1,2,3$. In Fig.~1, we have singled out the direction associated with the couplings $J_1,\;J_1'$. We consider a large lattice with periodic boundary conditions along the axes associated with the parameters $J_{2,3}, \; J_{2,3}'$.
The corresponding matrix can be represented as follows:
\bea
\frac{1}{2}W=\qquad \qquad \qquad \qquad \qquad \hspace{10cm}\nn\\
\left(\ba{cccc}e^{J_{1+1'}}\cosh{[J_{2+2'+3+3'}]}&e^{J_{1'-1}}\cosh{[J_{2+2'+3-3'}]}
&e^{J_{1-1'}}\cosh{[J_{2-2'+3+3'}]}&e^{J_{-1'-1}}\cosh{[J_{2-2'+3-3'}]}\\
e^{J_{1-1'}}\cosh{[J_{2+2'-3+3'}]}&e^{J_{-1'-1}}\cosh{[J_{2+2'-3-3'}]}
&e^{J_{1+1'}}\cosh{[J_{2-2'-3+3'}]}&e^{J_{-1'+1}}\cosh{[J_{2-2'-3-3'}]}\\
e^{J_{1'-1}} \cosh{[J_{2-2'-3-3'}]}
&e^{J_{1'+1}}\cosh{[J_{2-2'-3+3'}]}
&e^{J_{-1-1'}}\cosh{[J_{2+2'-3-3'}]}&e^{J_{1-1'}}\cosh{[J_{2+2'-3+3'}]}\\
e^{J_{-1-1'}}\cosh{[J_{2-2'+3-3'}]}&e^{J_{1-1'}}\cosh{[J_{2-2'+3+3'}]}
&e^{J_{1'-1}}\cosh{[J_{2+2'+3-3'}]}&e^{J_{1+1'}}\cosh{[J_{2+2'+3+3'}]}\ea\right)&
\nn
\ena
Here we use the notation $J_{k\pm p'\pm...}=J_k\pm J_p'\pm...$.

For the three-parameter case $J'_k=J_k,\quad k=1,2,3$, the weight matrix takes the form
\bea
2\left(\ba{cccc}e^{2J_1}\cosh{2[J_2+J_3]}&\cosh{2J_2}&\cosh{2J_3} & e^{-2J_1}\\
\cosh{2J_2} & e^{-2J_1}\cosh{2[J_2-J_3]} & e^{2J_1}&\cosh{2J_3} \\cosh{2J_3}
&e^{2J_1}& e^{-2J_1}\cosh{2[J_2-J_3]}&\cosh{2J_2}\\
e^{-2J_1}&\cosh{2J_3}&\cosh{2J_2}&e^{2J_1}\cosh{2[J_2+J_3]}\ea\right)\quad
\ena
Then, as in \cite{KhS1}, the partition function can be written in terms of a product of transfer matrices defined by
\bea
Z=Tr_{\{s_{(2i+1,1)}\}_{\; i=1,...,N}} \prod_j\tau_j,\quad \tau_j=Tr_{s_{(0,2j)}}
	\prod_i R_{(2i,2j)(2i+1,2j-1)}^{(2i+2,2j)(2i+1,2j+1)}, \label{trR}
\ena
Here we use the index convention $i=0,...,N-1$, $j=1,...,N$, and impose the following periodic boundary conditions on the model: $s_{p,k+N}=s_{p,k}$, $s_{p+N,k}=s_{p,k}$.

We note that in the earlier work \cite{Kagome1}, a different method was used to evaluate the partition function of the homogeneous model. The advantage of the representation (\ref{partw}, \ref{trR}) is that it allows us to formulate the general model as an eight-vertex model with an inhomogeneous $R$-matrix, which coincides with the weight function $W_{ij}$ up to a local unitary transformation, as was done in our work \cite{KhS1}. By attaching to each vertex the identity operator $I=U U^{-1}$, with a unitary operator
\bea
U=\frac{1}{\sqrt{2}}\left(\ba{cc} 1&-1\\
1&1\ea\right),
\ena
we can rewrite the partition function as
\bea
Z=\sum_{s_{\alpha\beta}}\prod_{\alpha\beta}R_{\alpha\beta},\quad R_{\alpha\beta}=U^{-1}_\alpha U^{-1}_\beta W_{\alpha\beta} U_{\alpha'}U_{\beta'}. \label{partr}
\ena
The $R$-matrix has the standard form of the eight-vertex model
\bea R=\left(\ba{cccc}R_{00}^{00}&0&0&R_{00}^{11}\\
0&R_{01}^{01}&R_{01}^{10}&0\\
0&R_{10}^{01}&R_{10}^{10}&0\\
R_{11}^{00}&0&0&R_{11}^{11}
\ea\right)\ena
with the following matrix elements:
\bea \label{R1}
R_{00}^{00}=8\left(\cosh{J_1'}\cosh{J_2'}\cosh{J_3}+\sinh{J_1'}\sinh{J_2'}\sinh{J_3}\right)\times\nn\\
\left(\cosh{J_1}\cosh{J_2}\cosh{J_3'}+\sinh{J_1}\sinh{J_2}\sinh{J_3'}\right),\nn\\
R_{11}^{11}=8\left(\sinh{J_1'}\cosh{J_2'}\cosh{J_3}+\cosh{J_1'}\sinh{J_2'}\sinh{J_3}\right)\times\nn\\\nn
\left(\sinh{J_1}\cosh{J_2}\cosh{J_3'}+\cosh{J_1}\sinh{J_2}\sinh{J_3'}\right),\\\nn
R_{00}^{11}=8\left(\cosh{J_1'}\sinh{J_2'}\cosh{J_3}+\sinh{J_1'}\cosh{J_2'}\sinh{J_3}\right)\times\\\nn
\left(\cosh{J_1}\cosh{J_2}\sinh{J_3'}+\sinh{J_1}\sinh{J_2}\cosh{J_3'}\right),\\\nn
R_{11}^{00}=8\left(\cosh{J_1'}\cosh{J_2'}\sinh{J_3}+\sinh{J_1'}\sinh{J_2'}\cosh{J_3}\right)\times\\\nn
\left(\cosh{J_1}\sinh{J_2}\cosh{J_3'}+\sinh{J_1}\cosh{J_2}\sinh{J_3'}\right),\\\nn
R_{01}^{10}=8\left(\sinh{J_1'}\cosh{J_2'}\cosh{J_3}+\cosh{J_1'}\sinh{J_2'}\sinh{J_3}\right)\times\\
\left(\cosh{J_1}\cosh{J_2}\cosh{J_3'}+\sinh{J_1}\sinh{J_2}\sinh{J_3'}\right),\\\nn
R_{10}^{01}=8\left(\cosh{J_1'}\cosh{J_2'}\cosh{J_3}+\sinh{J_1'}\sinh{J_2'}\sinh{J_3}\right)\times\\\nn
\left(\sinh{J_1}\cosh{J_2}\cosh{J_3'}+\cosh{J_1}\sinh{J_2}\sinh{J_3'}\right),\\\nn
R_{01}^{01}=8\left(\sinh{J_3}\cosh{J_2'}\cosh{J_1'}+\cosh{J_3}\sinh{J_2'}\sinh{J_1'}\right)\times\\\nn
\left(\sinh{J_3'}\cosh{J_2}\cosh{J_1}+\cosh{J_3'}\sinh{J_2}\sinh{J_1}\right),\\\nn
R_{10}^{10}=8\left(\sinh{J_2'}\cosh{J_1'}\cosh{J_3}+\cosh{J_2'}\sinh{J_1'}\sinh{J_3}\right)\times\\\nn
\left(\sinh{J_2}\cosh{J_1}\cosh{J_3'}+\cosh{J_2}\sinh{J_1}\sinh{J_3'}\right).
\ena
%

%
%
%

As a first consistency check, we verify the free-fermion condition characteristic of the 2D Ising model. It is fulfilled in the present case as well, for arbitrary $J,J'$:
\bea
R_{00}^{00}R_{11}^{11}-R_{00}^{11}R_{11}^{00}=R_{01}^{10}R_{10}^{01}-
R_{01}^{01}R_{10}^{10}.
\ena
This allows us to represent the partition function in a fermionic form with a quadratic fermionic action, writing it as a functional integral over Grassmann variables, as was done in Ref.~\cite{FerKagome}, and obtaining the corresponding expression as a product of determinants of the fermionic action
\bea
Z=tr \prod^{N\times N}R_{i,k}=[R_{00}^{00}]^{2N^2}\prod_{p,q}Det[\mathcal{A}(p,q)]
\ena
and explore the physical characteristics, say as critical points, thermal capacity (see the next sections) and so on.  Alternatively, the partition function can be obtained using the Kac--Ward representation; see, for example, Ref.~\cite{KWK}.

As in Ref.~\cite{KhS1}, we shall consider periodic boundary conditions. The partition function can then be represented as a product over the momentum-space variables:
\bea
\label{Zj}
Z=\prod_{n_1=1}^{N/2}\prod_{n_2=1}^{N}
\Bigl[
\mathbb{A}_1(J_k,J'_k)
+\mathbb{A}_2(J_k,J'_k)\cos[p]
+\mathbb{A}_3(J_k,J'_k)\cos[q]
+\mathbb{A}_4(J_k,J'k)\cos[p+q]
\Bigr],\;\quad
\ena
where
$ p=\frac{\pi(2n_1+1)}{N},
\;\;
q=\frac{\pi(2n_2+1)}{N}.$
The coefficients entering Eq.~(\ref{Zj}) are
\bea
\label{A1}
\mathbb{A}_1(J_k,J'_k)
&=&
2^4\sum_{k=1}^{3}
\cosh[2J_k]\cosh[2J'k]
\\
&+&2^4
\left(
\prod_{k=1}^{3}\cosh[2J_k]
+\prod_{k=1}^{3}\sinh[2J_k]
\right)
\left(
\prod_{k=1}^{3} \cosh[2J'k]
+\prod_{k=1}^{3} \sinh[2J'_k]
\right),
\nn\\[2mm]
\label{A2}
\mathbb{A}_2(J_k,J'_k)
&=&
-2^4
\Bigl(
\cosh[2J_1]\sinh[2J'_1]\sinh[2J_2]\sinh[2J_3]
\\
&&\hspace{1.2cm}
+\sinh[2J_1]\cosh[2J'_1]\sinh[2J'_2]\sinh[2J'_3]
\Bigr)
\nn\\
&-&2^4\sinh[2J_1]\sinh[2J'_1]
\Bigl(
\cosh[2J_2]\cosh[2J_3]
+\cosh[2J'_2]\cosh[2J'_3]
\Bigr),
\nn\\[2mm]
\label{A3}
\mathbb{A}_3(J_k,J'_k)
&=&
-2^4
\Bigl(
\cosh[2J_3]\sinh[2J'_3]\sinh[2J_2]\sinh[2J_1]
\\
&&\hspace{1.2cm}
+\sinh[2J_3]\cosh[2J'_3]\sinh[2J'_2]\sinh[2J'_1]
\Bigr)
\nn\\
&&
-2^4\sinh[2J_3]\sinh[2J'_3]
\Bigl(
\cosh[2J_2]\cosh[2J_1]
+\cosh[2J'_2]\cosh[2J'_1]
\Bigr),
\nn\\[2mm]
\label{A4}
\mathbb{A}_4(J_k,J'_k)
&=&
-2^4
\Bigl(
\cosh[2J_2]\sinh[2J'_2]\sinh[2J_1]\sinh[2J_3]
\\
&&\hspace{1.2cm}
+\sinh[2J_2]\cosh[2J'_2]\sinh[2J'_1]\sinh[2J'_3]
\Bigr)
\nn\\
&&
-2^4\sinh[2J_2]\sinh[2J'_2]
\Bigl(
\cosh[2J_1]\cosh[2J_3]
+\cosh[2J'_1]\cosh[2J'_3]
\Bigr).\nn
\ena
As can be seen, the functions $\mathbb{A}_i(J_k,J'_k)$, $i=2,3,4$, are related by permutations of the three pairs of coupling constants
${J_k,J'_k}$, $k=1,2,3$. Defining
\bea
\mathbb{A}
\left(
{J_1,J'_1},
{J_2,J'_2},
{J_3,J'_3}
\right)
\equiv
\mathbb{A}_2(J_k,J'_k),
\ena
one obtains
\bea
\mathbb{A}_3(J_k,J'_k)
=
\mathbb{A}
\left(
{J_3,J'_3},
{J_1,J'_1},
{J_2,J'_2}
\right), \qquad
\mathbb{A}_4(J_k,J'_k)
=
\mathbb{A}
\left(
{J_2,J'_2},
{J_3,J'_3},
{J_1,J'_1}
\right).\qquad
\ena
Thus, $\mathbb{A}_2$, $\mathbb{A}_3$, and $\mathbb{A}_4$ are generated from one another by cyclic permutations of the three lattice directions. The same permutation structure was also encountered in Ref.~\cite{FerKagome}.

 \section{Yang-Baxter equations, integrability}

 In the work \cite{KhS2} we have analyzed the the Yang-Baxter equations (YBE) with the general eight-vertex form. There we have obtained rather general
 conditions,  for satisfying the spectral parameter dependent  YBE,
 \bea
 R_{12}(\mathbf{u},\mathbf{v})
R_{13}(\mathbf{u},\mathbf{w})R_{23}(\mathbf{v},\mathbf{w})=
R_{23}(\mathbf{v},\mathbf{w})R_{13}(\mathbf{u},\mathbf{w})R_{12}(\mathbf{u},\mathbf{v}),
\label{ybe} \ena
where the spectral parameters $\mathbf{u},\mathbf{v},\mathbf{w}$ are in general complex variables, which are connected with the model parameters. 
Colored solutions are permissible if the free fermionic condition is fulfilled. Here, for simplicity we shall consider two cases: ${J'}_k=\pm J_k$.


The $R$-matrix for the case $J'_k=J_k$ can be presented in this normalised form:

\bea
R^+_{J_k}=\left(\ba{cccc}
F[J_1;J_2,J_3]&0&0&\frac{F[J_3;J_1,J_2]F[J_2;J_1,J_3]}{F[J_1;J_2,J_3]}\\
0&\frac{F[J_2;J_1,J_3]^2}{F[J_1;J_2,J_3]}&1&0\\
0&1&\frac{F[J_3;J_1,J_2]^2}{F[J_1;J_2,J_3]}&0\\
\frac{F[J_2;J_1,J_3]F[J_3;J_1,J_2]}{F[J_1;J_2,J_3]}&0&0&\frac{1}{F[J_1;J_2,J_3]}\ea
\right)
\ena
where
\bea
F[J_k;J_r,J_p]=\frac{\tanh[J_k]+\tanh[J_r]\tanh[J_p]}{1+\tanh[J_k]\tanh[J_r]\tanh[J_p]}.
\ena

For the sign-altered case we have
\bea
R^-_{J_k}=\left(\ba{cccc}
F[J_1;J_2,J_3]&0&0&\frac{-F[J_3;J_1,J_2]G[J_2;J_1,J_3]}{G[J_1;J_2,J_3]}\\
0&\frac{F[J_2;J_1,J_3]G[J_2;J_1,J_3]}{G[J_1;J_2,J_3]}&-1&0\\
0&\frac{F[J_1;J_2,J_3]}{G[J_1;J_2,J_3]}&\frac{F[J_3;J_1,J_2]G[J_3;J_1,J_2]}{G[J_1;J_2,J_3]}&0\\
\frac{-F[J_2;J_1,J_3]G[J_3;J_1,J_2]}{G[J_1;J_2,J_3]}&0&0&\frac{-1}{G[J_1;J_2,J_3]}\ea
\right)
\ena
where
\bea
G[J_1;J_2,J_3]=\frac{\tanh[J_k]-\tanh[J_r]\tanh[J_p]}{1-\tanh[J_k]\tanh[J_r]\tanh[J_p]}.
\ena

For this case we deal with a graded matrix. For escaping to introduce the parities of the states in the YB equations, we must consider the check matrices: $\check{R}_{ab}^{a'b'}=R_{ab}^{b'a'}$, as in this formulation the check YBE do not carry additional signs
\bea
\check{R}_{12}(\mathbf{u},\mathbf{v})
\check{R}_{23}(\mathbf{u},\mathbf{w})\check{R}_{12}(\mathbf{v},\mathbf{w})=
\check{R}_{23}(\mathbf{v},\mathbf{w})\check{R}_{12}(\mathbf{u},\mathbf{w})\check{R}_{23}(\mathbf{u},\mathbf{v})\ena
Here the arguments $(\mathbf{u},\mathbf{v})$ in general are complex variables, which must be connected by the statistical system parameters $J_k,\; k=1,2,3$
by some  elliptic functions \cite{KhS1,KhYBE}.

One can easily verify that the matrix $R^+(J_k)$ at the limit $\tanh[J_1]=1$, $(\tanh[J_2+J_3])^2=\tanh[u-w]$, which corresponds to the operator form
\bea\check{R}(u,w)=I\otimes I +\tanh[u-w]\sigma_1\otimes \sigma_1,\label{Rpm1}\ena
the $R(u,w)$-matrix
satisfies the ordinary  one-parametric Yang-Baxter relations:\\ $R_{12}(u-w)R_{13}(u-v)R_{23}(v-w)=R_{23}(v-w)R_{13}(u-v)R_{12}(u-w)$.

The similar limit for the matrix $R^-(J_k)$ appears to be $\tanh[J_1]=1$, $\tanh[J_2+J_3]\tanh[J_2-J_3]=\tanh[u-w]$, satisfying to the graded one-parametric YBE or for
the check form, which has the same operator form (\ref{Rpm1}), the ordinary non-graded YBE: \\
$\check{R}_{12}(u-w)\check{R}_{23}(u-v)\check{R}_{12}(v-w)=\check{R}_{23}(v-w)\check{R}_{12}(u-v)\check{R}_{23}(u-w)$.

 \section{1D quantum model}
 \subsection{Ferromagnetic 1d model}

For the simple anisotropic model $J_k={J'}_k$, let us consider continuum limit in order to obtain the exact Hamiltonian of the corresponding one-dimensional quantum spin chain. Taking $e^{-J_1}=h t$, $J_2=J_p t$, and $J_3=J_q t$, with a small parameter $t\ll 1$, the normalized $\check{R}(J_1,J_2,J_3)$ operator
\bea
\check{R}(J_1,J_2,J_3)=\frac{1}{8(\cosh[J_1]\cosh[J_2]\cosh[J_3])^2}P R(J_1,J_2,J_3)
\ena
admits such expansion over the parameter $t$:
\bea
&&\check{R}(h,J_{p},J_{q})=\check{R}(J_1,J_2,J_3)=\\\nn
&&\left(\ba{cccc}1+2 J_p J_q t^2&0&0&(J_p+J_q)^2 t^2\\
0&1+2 (J_p J_q-h^2)t^2&(J_p+J_q)^2 t^2&0\\
0&(J_p+J_q)^2 t^2&1+2(J_p J_q-h^2)t^2&0\\
(J_p+J_q)^2 t^2&0&0&1+2(J_p J_q-2 h^2)t^2
\ea\right)
\ena
In the operator form this matrix reads as:
\bea
\check{R}(h,J_{p},J_{q})=I\otimes I+2 \big(J_p J_q -h^2\big)t^2 I \otimes I+h^2 t^2 \big(I\otimes \sigma_z+\sigma_z\otimes I\big)+(J_p+J_q)^2 t^2(\sigma_x\otimes \sigma_x).\nn
\ena
Using the Bethe ansatz technique and treating the parameter $t^2$ as the spectral parameter, we obtain a one-dimensional chain described by the following IM Hamiltonian, up to the constant term $\sum_i (2 J_p J_q-2 h^2)t^2 =2N (J_p J_q-h^2)t^2$:
\bea
H=-\sum_{i} (2h^2 \sigma^z_{i}+(J_p+J_q)^2 \sigma^x_{i}\otimes \sigma^x_{i+1})
\ena
This ferromagnetic model undergoes a continuous phase transition at $2h^2=(J_p+J_q)^2$, which is consistent with the critical-surface condition of the anisotropic kagom\'e lattice,
\bea
\sum_{k}\cosh[2J_k]-\prod_k \sinh[2J_k]-\prod_k \cosh[2J_k]=0
\ena
in the limit $t\to 0$, after applying the transformations introduced above.

\subsection{ Sign-altered anisotropic  1d model}

Here we consider the partially antiferromagnetic choice of the parameters ${J'}_k=-J_k$. In this case, we obtain the following $R_s$ and $\check{R_s}$ matrices, using the notation $z_k=\tanh{[J_k]}$.
\bea
&&{R_s}(J_1,J_2,J_3)=\frac{1}{8(\cosh[J_1]\cosh[J_2]\cosh[J_3])^2}=\qquad\\\nn
&&{\scriptsize{\left(\ba{cccc}
1-z_1^2 z_2^2 z_3^2&0&0&(z_2+{z_1}{z_3})({z_3}-{z_1}{z_2})\\
0&z_1^2 z_2^2-z_3^2&(z_1+{z_2}{z_3})({z_1}{z_2}z_3-1)&0\\
0&(z_1-{z_2}{z_3})({z_1}{z_2}{z_3}+1)&z_1^2 z_3^2-z_2^2&0\\
(z_2-{z_1}{z_3})(z_3+{z_1}{z_2})&0&0&z_2^2 z_3^2-z_1^2\ea
\right)}}
\ena

A smoother version is:

In the limit $e^{-J_1}=h t$, $J_2=J_p t$, and $J_3=J_q t$, with a small parameter $t$, we obtain:
\bea
&&\check{R_s}(h,J_{p},J_{q})=R_s((h,J_{p},J_{q})\times [R_s((h,J_{p},J_{q})_{t\to 0}]^{-1}\\\label{pgrad}
&=&R_s(h,J_{p},J_{q})\times \left(\ba{cccc}1&0&0&0\\
0&0&1&0\\
0&-1&0&0\\
0&0&0&-1
\ea\right)\\\label{hgrad}
&=&\left(\ba{cccc}1&0&0&(J_p^2-J_q^2) t^2\\
0&1-2 h^2 t^2&(J_p^2-J_q^2) t^2&0\\
0&(J_p^2-J_q^2) t^2&1-2 h^2 t^2&0\\
(J_p^2-J_q^2) t^2&0&0&1-4 h^2 t^2
\ea\right)
\ena
In this case, instead of the ordinary permutation matrix, we obtain the graded permutation matrix $P_s=[R_s((h,J_p,J_q)_{t\to0}]^{-1}$, as presented in Eq.~(\ref{pgrad}). It follows from the matrix in Eq.~(\ref{hgrad}) that the corresponding one-dimensional quantum IM Hamiltonian is
\bea
H_s=-\sum_{i} (2h^2 {{\sigma_z}}_{i}+(J_p^2-J_q^2)[{\sigma_x}_{i}\otimes {\sigma_x}_{i+1}])
\ena
In the ferromagnetic case $(J_p^2-J_q^2)>0$ the model admits continuous phase transition at $2h^2=J_p^2-J_q^2$.

The partition function of this model is
\bea
Z&=&Z_0\prod_{p,q}\left(-8 \tanh[J_1]^2 \tanh[J_2]^2 \tanh[
J_3]^2 + (1 + \tanh[J_1]^4) (1 + \tanh[J_2]^4) (1 + \tanh[J_3]^4)\right.\nn\\
&+&\left. 2 \cosh[2 J_2] \sec[J_2]^4 (\cos[2 p] \cosh[2 J_3] \sec[J_3]^4
\tanh[J_1]^2 \right.\nn\\
&+&\left. \cos[2 q] \cosh[2 J_1] \sec[J_1]^4 \tanh[J_3]^2) \right.\\
&+& \left. 2 \cos[2 (p - q)] \cosh[2 J_1] \cosh[2 J_3] \sec[J_1]^4 \sec[J_3]^4 \tanh[J_2]^2 \right)\nn
\ena
and admits zeros at the following points of the momentum spectrum: ${p,q}={\pi/2,\pi/2}$, ${0,\pi/2}$, and ${\pi/2,0}$ (up to ${\pm \pi,\pm \pi}$), whereas the previous case corresponds to the usual choice of zero momenta, $p=q=0$. In particular, here we discuss in detail the case ${p,q}={\pi/2,\pi/2}$, since the other two cases are related to it by permutations of the parameters $J_k$: ${J_1,J_2,J_3}$ $\to$ ${J_3,J_1,J_2}$, ${J_2,J_3,J_1}$.

The critical surface at ${p,q}={\pi/2,\pi/2}$ is determined by the relation:
\bea
\left(1-\sum_{k\neq r}(-1)^{[k+r]}(\cosh[2J_k]\cosh[2J_r])\right)=0.
\ena

This equation for the critical surface leads to the following graphical representation. In the first (left-hand) plot, the parameter $J_3$ is shown along the vertical axis (and may equivalently be replaced by $J_1$), while the horizontal axes correspond to the parameters $J_1$ (or $J_3$) and $J_2$:
\bea
J_{3/1}=\pm\frac{ 1}{2} \mathrm{Arctanh}[\frac{1+\cosh[2J_{1/3}]\cosh[2J_2]}{\cosh[2J_{1/3}]-\cosh[2J_2]}]
\ena
while in the second (right-hand) plot the parameter $J_2$ is shown along the vertical axis:
\bea
J_2=\pm\frac{ 1}{2} \mathrm{Arctanh}[\frac{-1+\cosh[2J_1]\cosh[2J_3]}{\cosh[2J_1]+\cosh[2J_3]}]
\ena
\begin{figure}[ht]\unitlength=8pt
	\begin{picture}(100,20)(-5,0)
		\includegraphics[width=70mm]{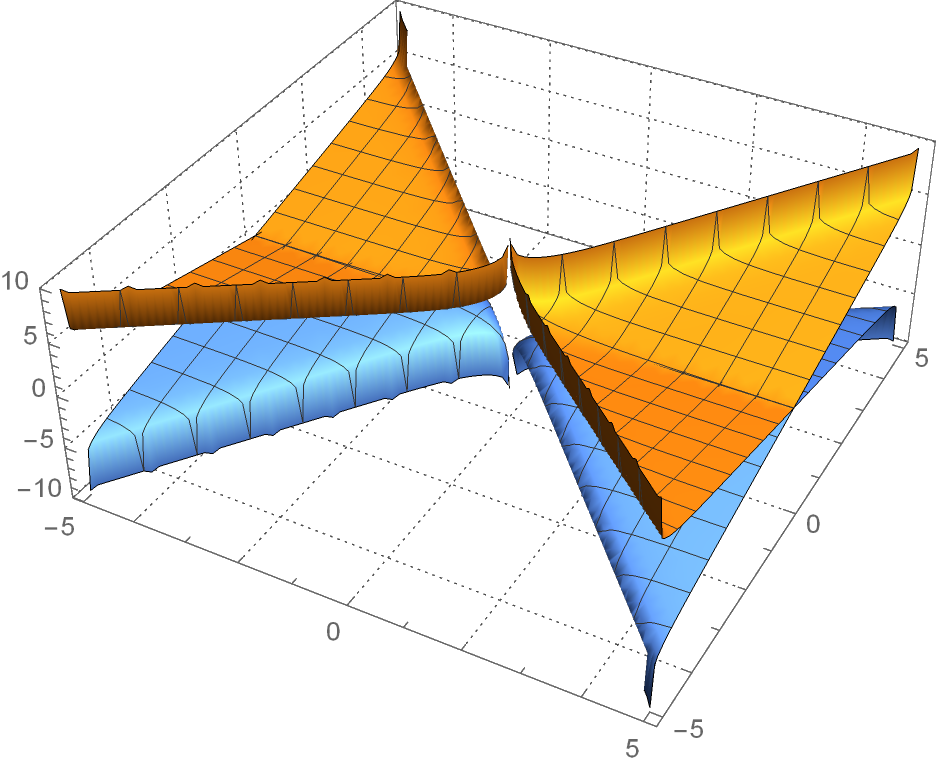}
		\includegraphics[width=70mm]{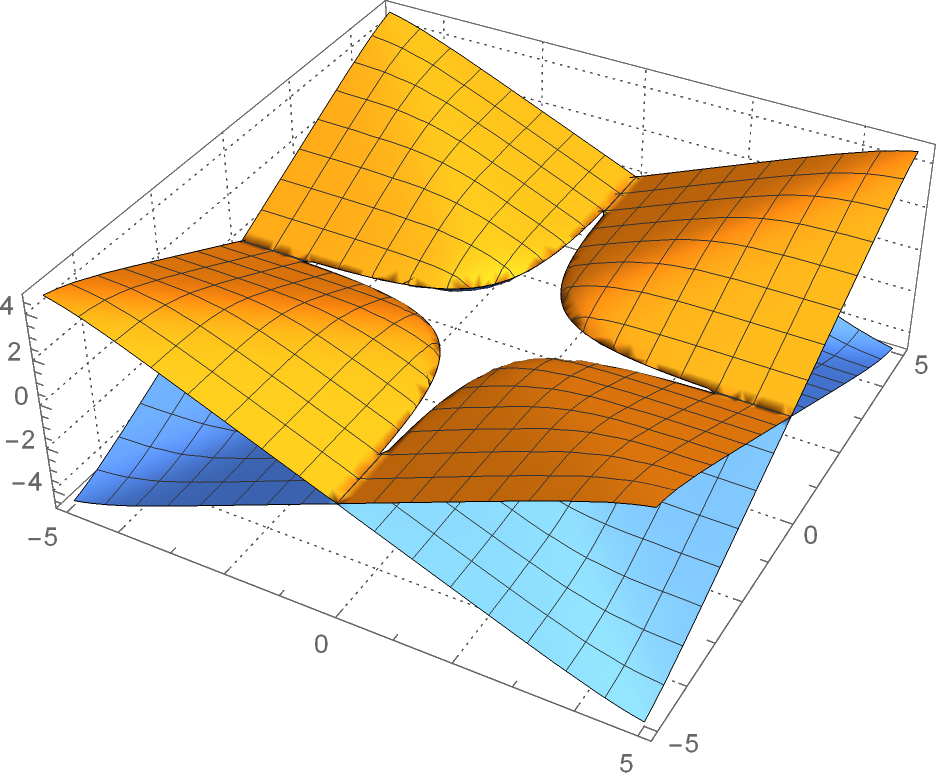}
	\end{picture}
	\caption{Critical subsurfaces on the real coordinate space {$J_1,J_2,J_3$}}
	\label{Surf1}
\end{figure}

The graphical representation of the critical surface for the choice ${J'}_k=J_k$ was presented in Ref.~\cite{FerKagome}. In that case, the corresponding relation is invariant under permutations of all parameters  $J_k$.

\section{Specific heat calculation for the partly anisotropic case: $J_2=J_3=J$, ${J'}_2={J'}_3=J'$}


From the expression for the partition function (\ref{Zj}), in the particular limit of the coupling constants
\bea
J_2=J_3=J,\quad {J'}_2={J'}_3=J',\quad J_1=\bar{J},\quad J'_1=\bar{J}',
\ena
we obtain the following expression for the free energy per site,
$F=-T\ln[Z]/N^2$, in the thermodynamic limit:
\bea\label{fjjx}
&F\left(\frac{J}{T},\frac{J'}{T},\frac{\bar{J}}{T},\frac{\bar{J}'}{T}\right)=&\\ \nn
&-\frac{T}{\pi^2}\int_{0}^{\frac{\pi}{2}}dp\int_0^{\pi} dq
\ln\left[\mathbb{A}_x(T)-\mathbb{A}_y
\left(T\right)\cos[2p]-\mathbb{A}_z(T)\left(\cos[2q]+\cos[2p+2q]\right)\right].&
\ena
The functions $\mathbb{A}_{x,y,z}(T)$ are obtained from the coefficients
$\mathbb{A}_{1,2,3,4}\left(J_k/T,J'_k/T\right)$ introduced in Section~2.
For the present choice of parameters
$J,\;J',\;\bar{J},\;\bar{J}'$, they take the form
\bea
&\mathbb{A}_x(T)=2^{-4}\mathbb{A}_1\left(\frac{J}{T},\frac{J'}{T},\frac{\bar{J}}{T},\frac{\bar{J}'}{T}\right)=
2\cosh[\frac{2J}{T}]\cosh[\frac{2J'}{T}]+
\cosh[\frac{2\bar {J}}{T}]\cosh[\frac{2\bar{J}'}{T}]+&\nn\\
& (\cosh[\frac{2J}{T}]^2\cosh[\frac{2\bar{J}}{T}]+\sinh[\frac{2J}{T}]^2\sinh[\frac{2\bar{J}}{T}])
(\cosh[\frac{2J'}{T}]^2\cosh[\frac{2\bar{J}'}{T}]+\sinh[\frac{2J'}{T}]^2\sinh[\frac{2\bar{J}'}{T}]),&\\\nn
& \mathbb{A}_y(T)=-2^{-4}\mathbb{A}_2\left(\frac{J}{T},\frac{J'}{T},\frac{\bar{J}}{T},\frac{\bar{J}'}{T}\right)=
\sinh[\frac{2J}{T}]^2\cosh[\frac{2\bar{J}}{T}]\sinh[\frac{2\bar{J}'}{T}]+\sinh[\frac{2{J}'}{T}]^2\sinh[\frac{2\bar{J}}{T}]\cosh[\frac{2\bar{J'}}{T}]+&\\&
\sinh[\frac{2\bar{J}'}{T}]\sinh[\frac{2\bar{J}}{T}](\cosh[\frac{2J}{T}]^2+\cosh[\frac{2J'}{T}]^2)&\\&\nn
\mathbb{A}_z(T)=-2^{-4}\mathbb{A}_{3,4}\left(\frac{J}{T},\frac{J'}{T},\frac{\bar{J}}{T},\frac{\bar{J}'}{T}\right)=
\frac{1}{2}\left(\sinh[\frac{4J}{T}]\sinh[\frac{2\bar{J}}{T}]\sinh[\frac{2{J}'}{T}]+\sinh[\frac{4{J}'}{T}]\sinh[\frac{2{J}}{T}]\sinh[\frac{2\bar{J'}}{T}]\right)+&\\
&\sinh[\frac{2{J}'}{T}]\sinh[\frac{2{J}}{T}](\cosh[\frac{2\bar{J}}{T}]\cosh[\frac{2J}{T}]+\cosh[\frac{2J'}{T}]\cosh[\frac{2\bar{J}'}{T}])&
\ena

In the simpler case $J'=J$ and $\bar{J}'=\bar{J}$, considered in Ref.~\cite{FerKagome}, these expressions reduce to
\bea
\nn
&&\mathbb{A}_1=4\Big(6+6\cosh\Big[\frac{4J}{T}\Big]+3\cosh\Big[\frac{4\bar{J}}{T}\Big]+{\cosh\Big[\frac{4J}{T}\Big]}^2\cosh\Big[\frac{4\bar{J}}{T}\Big]+{\sinh\Big[\frac{4J}{T}\Big]}^2\sinh{\Big[\frac{4\bar{J}}{T}\Big]}\Big),\\
&&\mathbb{A}_{3,4}=\mathbb{A}\left(\frac{J}{T},\frac{J}{T},\frac{\bar{J}}{T}\right)=\mathbb{A}\left(\frac{\bar{J}}{T},\frac{J}{T},\frac{J}{T}\right)=-16 \sinh\Big[\frac{2J}{T}\Big]\sinh\Big[\frac{4J}{T}\Big]e^{\frac{2\bar{J}}{T}},\\ \nn
&&\mathbb{A}_2=\mathbb{A}\left(\frac{J}{T},\frac{\bar{J}}{T},\frac{J}{T}\right)=-8\left(\cosh\Big[\frac{4J}{T}\Big]+1\right)\left(\cosh\Big[\frac{4\bar{J}}{T}\Big]-1\right)
-8\sinh\Big[\frac{4\bar{J}}{T}\Big]\left(\cosh\Big[\frac{4J}{T}\Big]-1\right).
\ena

The specific heat is defined by
\bea\label{heat}
C=-T\frac{\partial^2 F(T)}{\partial T^2}.
\ena
For the fully isotropic case, this quantity was calculated in Ref.~\cite{FerKagome}. Here we extend the calculation to the partially anisotropic case, for which the free energy retains the general integral structure of Eq.~(\ref{fjjx}). Differentiating Eq.~(\ref{fjjx}) twice with respect to temperature gives the following integral representation for the specific heat:
\bea
\label{sheat}
&C=-\frac{T}{\pi^2}\int_{0}^{\frac{\pi}{2}}dp\int_0^{\pi} dq\left(
T \frac{(\mathbb{A}'_x({T})-\mathbb{A}'_y({T})\cos[2p]-2\mathbb{A}'_z
	({T})\cos[2q+p]\cos[p])^2}{(\mathbb{A}_x({T})-\mathbb{A}_y({T})\cos[2p]-2\mathbb{A}_z({T}) \cos[2q+p]\cos[p])^2}\right.&
\\&-
\left.\frac{2(\mathbb{A}'_x({T})-\mathbb{A}'_y({T})\cos[2p]-2\mathbb{A}'_z
	({T})\cos[2q+p]\cos[p])+T(\mathbb{A}''_x({T})-\mathbb{A}''_y({T})\cos[2p]-2\mathbb{A}''_z
	({T})\cos[2q+p]\cos[p])}{\mathbb{A}_x({T})-\mathbb{A}_y({T})\cos[2p]-2\mathbb{A}_z({T}) \cos[2q+p]\cos[p]}\right)\nn&
\ena

Here and below, we use the conventional notation
\bea
\mathbb{A}'_a(T)
=\frac{\partial\mathbb{A}_a(T)}{\partial T}
=-
\sum_k \frac{J_k}{T^2}
\frac{\partial \mathbb{A}_a (\ldots,J_k/T,\ldots) }
{\partial(J_k/T)},
\ena
with an analogous definition for $\mathbb{A}_a(T)$.  Two parameter integral expression  (\ref{sheat}) of the specific heat can be reduced to one parameter integral
\bea
\label{C00}
C=C_0(T)+\frac{T}{\pi}\int_0^{\pi/2}dp\frac{2Q[A_{x,y,z}(T),\cos[2p]]}{\sqrt{(A_x(T)-A_y(T)\cos[2p])^2-4\cos[p]^2 A_z(T)^2}}+\nn\\
\nn\frac{T}{\pi}\int_0^{\pi/2}dp\frac{2W[A_{x,y,z}(T),\cos[2p]](A_x(T)-A_y(T)\cos[2p])}{\left[(A_x(T)-A_y(T)\cos[2p])^2-4\cos[p]^2 A_z(T)^2\right]^{3/2}}\\
C_0(T)=\frac{T}{2}\left( \frac{2\mathbb{A}'_z(T)}{\mathbb{A}_z(T)} -T \left[\frac{(\mathbb{A}_z(T))}{\mathbb{A}_z(T)}\right]^2
+\frac{T \mathbb{A}''_z(T)}{\mathbb{A}_z(T)}\right)=T \frac{\partial^2 (T \log{A_z(T)})}{2\partial T^2},
\ena
where the polynomial functions $Q,\;W$ can be presented in a rather compact form
\bea
Q[\mathbb{A}_{x,y,z}(T),\cos[2p]]=\mathbb{A}_z(T)\left(\frac{\partial^2[T \mathbb{A}_x(T)/\mathbb{A}_z(T)]}{\partial T^2} -
\cos[2p] \frac{\partial^2[T \mathbb{A}_(T)/\mathbb{A}_z(T)]}{\partial T^2}\right),\\
W[\mathbb{A}_{x,y,z}(T),\cos[2p]]=-T\mathbb{A}_z(T)^2 \left(\frac{\partial[\mathbb{A}_x(T)/\mathbb{A}_z(T)]}{\partial T} - \cos[2p]
\frac{\partial[\mathbb{A}_y(T)/\mathbb{A}_z[T]]}{\partial T}\right)^2
\ena
One can verify, that at the case $J'=J,\;\bar{J}'=\bar{J}$, the coefficient $C_0$ coincides with the isotropic case.


\section{Conclusion}

In this work we have developed an $R$-matrix formulation of a generalized
inhomogeneous Ising model on the two-dimensional kagom\'e lattice. The model
contains two alternating coupling constants $J_i$ and $J'_i$, $i=1,2,3$, along
the three lattice directions. By combining two neighboring triangles into an
elementary cell and performing a local unitary transformation, the local
Boltzmann weights are represented by a non-symmetric eight-vertex $R$-matrix.
A central property of this construction is that the resulting $R$-matrix
satisfies the free-fermion condition for arbitrary values of the six coupling
constants. This extends the connection between the Ising model, fermionic
representations, and eight-vertex models developed previously in
Ref.~\cite{KhS1}. The general Yang--Baxter structure of inhomogeneous
eight-vertex $R$-matrices, investigated in Ref.~\cite{KhS2}, provides the
algebraic framework for the analysis carried out here.

We have considered in detail the two choices
$J'_i=J_i$ and $J'_i=-J_i$. The first case contains the ordinary anisotropic
kagom\'e Ising model, whereas the second defines a sign-altered model with a
graded structure of the corresponding $R$-matrix. In the anisotropic, or
continuum, limit, the associated one-dimensional quantum Hamiltonians were
derived explicitly. For $J'_i=J_i$, the resulting Hamiltonian is of the
transverse-field Ising type, and its critical point agrees with the
corresponding anisotropic limit of the critical surface of the classical
kagom\'e model. For $J'_i=-J_i$, the limiting $R$-matrix involves a graded
permutation operator and leads to a modified Ising chain. In this case, the
zeros of the partition function occur at shifted momenta, in contrast with
the conventional ferromagnetic case.

For the partially anisotropic choice
$J_2=J_3=J$ and $J'_2=J'_3=J'$, we have also obtained the free energy in the
thermodynamic limit and derived the corresponding expression for the specific
heat. These results extend the thermodynamic analysis of the inhomogeneous
kagom\'e Ising model developed by the fermionic-field method in
Ref.~\cite{FerKagome}.

The present work therefore complements two earlier formulations of the same
class of models. In Ref.~\cite{FerKagome}, the inhomogeneous kagom\'e Ising
model was studied by means of a fermionic field representation, leading to
its thermodynamic functions and critical surface. In Ref.~\cite{KWK}, the
critical properties of the kagom\'e and other planar Ising lattices were
analyzed independently within the Kac--Ward formalism. The present
$R$-matrix approach adds the Yang--Baxter and quantum-spin-chain aspects to
these results. Taken together, the three approaches establish a coherent
connection between the geometry and thermodynamics of the two-dimensional
kagom\'e Ising model, its fermionic representation, its critical properties,
and the integrable structure of the associated one-dimensional quantum
models.


\section{Acknowledgments}
The authors  acknowledge support from the Armenian HESC grants 21AG-1C024  and 24FP-1F039 for financial support.

\end{document}